\documentclass[pdflatex,sn-mathphys-num]{sn-jnl}

\usepackage{graphicx}%
\usepackage{multirow}%
\usepackage{amsmath,amssymb,amsfonts}%
\usepackage{amsthm}%
\usepackage{mathrsfs}%
\usepackage[title]{appendix}%
\usepackage{xcolor}%
\usepackage{textcomp}%
\usepackage{manyfoot}%
\usepackage{booktabs}%
\usepackage{algorithm}%
\usepackage{algorithmicx}%
\usepackage{algpseudocode}%
\usepackage{listings}%
\begin{document}

\title[Article Title]{Promoting Cooperation in the Public Goods Game using Artificial Intelligent Agents}

\author*[1,2]{\fnm{Arend} \sur{Hintze}}\email{ahz@du.se}

\author[2,3,4,5]{\fnm{Christoph} \sur{Adami}}\email{adami@msu.edu}

\affil*[1]{\orgdiv{Department of MicroData Analytics}, \orgname{Dalarna University}, \orgaddress{\city{Falun}, \postcode{79188}, \country{Sweden}}}

\affil[2]{\orgname{BEACON Center for the Study of Evolution in Action}, \orgaddress{\city{East Lansing}, \state{Michigan}, \postcode{48824}, \country{USA}}}

\affil[3]{\orgdiv{Department of Microbiology, Genetics, and Immunology}, \orgname{Michigan State University}, \orgaddress{\city{East Lansing}, \state{Michigan}, \postcode{48824}, \country{USA}}}

\affil[4]{\orgdiv{Program in Ecology, Evolution, and Behavior}, \orgname{Michigan State University}, \orgaddress{\city{East Lansing}, \state{Michigan}, \postcode{48824}, \country{USA}}}

\affil[5]{\orgdiv{Department of Physics and Astronomy}, \orgname{Michigan State University}, \orgaddress{\city{East Lansing}, \state{Michigan}, \postcode{48824}, \country{USA}}}


\abstract{
The \textit{tragedy of the commons} illustrates a fundamental social dilemma where individual rational actions lead to collectively undesired outcomes, threatening the sustainability of shared resources. Strategies to escape this dilemma, however, are in short supply. In this study, we explore how artificial intelligence (AI) agents can be leveraged to enhance cooperation in public goods games, moving beyond traditional regulatory approaches to using AI as facilitators of cooperation. We investigate three scenarios: (1) \textit{Mandatory Cooperation Policy for AI Agents}, where AI agents are institutionally mandated always to cooperate; (2) \textit{Player-Controlled Agent Cooperation Policy}, where players evolve control over AI agents' likelihood to cooperate; and (3) \textit{Agents Mimic Players}, where AI agents copy the behavior of players. Using a computational evolutionary model with a population of agents playing public goods games, we find that only when AI agents mimic player behavior does the critical synergy threshold for cooperation decrease, effectively resolving the dilemma. This suggests that we can leverage AI to promote collective well-being in societal dilemmas by designing AI agents to mimic human players. 
}



\maketitle

\section{Introduction}\label{sec1}
The \textit{tragedy of the commons} \cite{hardin1968tragedy} refers to a fundamental social dilemma where individual rational actions lead to collectively undesired outcomes. In systems involving shared resources or public goods, individuals who cooperate by contributing are often disadvantaged compared to defectors who withhold their contributions. Because the public good is evenly distributed among all participants, defectors achieve higher personal gains than their cooperative counterparts because they reap the benefits without incurring the cost of contribution. This imbalance undermines cooperative behavior and threatens the shared resources' sustainability, posing significant challenges across disciplines such as economics, environmental science, and public policy.

\begin{figure}
    \centering
    \includegraphics[width=0.5\linewidth]{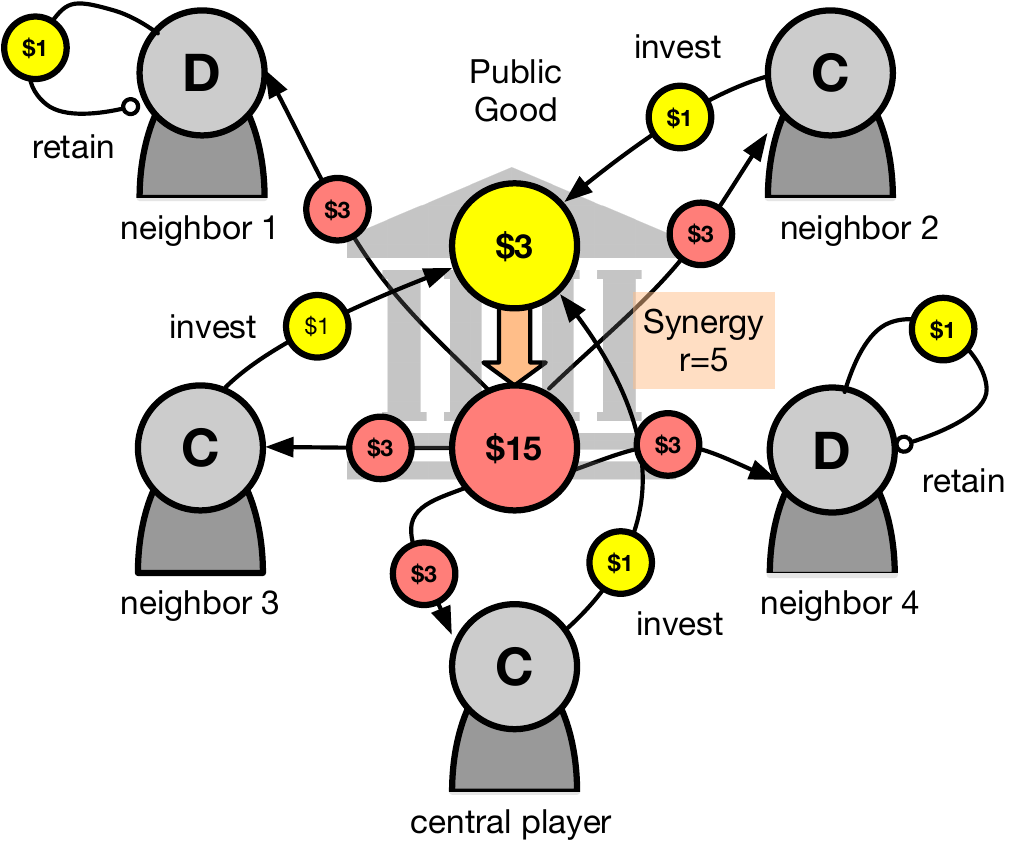}
    \caption{Illustration of the Public Goods Game. A central player and its ($k=4$) neighbors engaging in the public goods game. The three cooperators contribute \$1 to the public goods pool (invest), while two defectors withhold their \$1 investment (retain). A synergy ($r=5$) is applied which raises the value of the public goods pool (here \$15). The pool is redistributed equally to all players.}
    \label{fig:gameIllustration}
\end{figure}

The quintessential tool to study the emergence, evolution, and persistence of cooperation is Evolutionary Game Theory (EGT)~\cite{smith1982evolution,Axelrod1984,Nowak2006}. Within EGT, the {\em Public Goods} game~\cite{fehr2002altruistic} plays a prominent role: it is a multi-player extension of the well-known Prisoner's Dilemma game~\cite{Axelrod1984} that makes it possible to model the tragedy of the commons in terms of a {\em cooperation dilemma}.
At its core, the public goods game presents participants with a choice: to cooperate (C) by contributing a cost (or ante, here set to $1$ without loss of generality) to the common good, or to defect (D) by withholding that contribution. The total contributions are amplified by a multiplication factor $r$, symbolizing the {\em synergy} between cooperating parties. This enhanced public good is then evenly distributed among all participants, regardless of their contributions (see Figure~\ref{fig:gameIllustration}). In such a game, defectors always obtain higher rewards than cooperators in the same group, even though they could reap an even higher return if everyone cooperated. However, once the synergy factor $r$ becomes high enough ($r \ge k+1$, where $k+1$ is the group size), a defecting player fares worse than if they had cooperated instead. To promote cooperation, we ask: ``How can we change the game so that cooperation becomes the rational choice for synergies where $r<k+1$?''

Although numerous mechanisms have been proposed to promote cooperation in public goods dilemmas -- including the \textit{green beard effect}~\cite{hamilton1964evolution}, reciprocity \cite{nowak2006five}, spatial dynamics \cite{helbing2010punish}, group selection \cite{killingback2006evolution}, 
costly punishment ~\cite{fehr2000cooperation,fehr2002altruistic,hintze2015punishment,raihani2019punishment}, institutional incentives~\cite{Dongetal2016} and resource redistribution~\cite{hintze2020inclusive} -- these approaches do not account for the transformative impact of AI integration. Recognizing this gap, we pursue a novel exploration into how AI agents can be strategically leveraged to enhance cooperative behavior, an aspect previously unaddressed in the literature.

Governments are fundamentally responsible for establishing institutions that address the tragedy of the commons~\cite{ostrom1990governing}. By enacting legislation that incorporates mechanisms like costly punishment and resource redistribution, they create regulations, norms, and incentives that make cooperation more advantageous than defection~\cite{fehr2002altruistic,bowles2008policies,Dongetal2016}. For instance, legal penalties for free-riding discourage defection, while taxation and subsidies can redistribute resources to promote fairness~\cite{hardin1968tragedy}. By integrating these strategies, governments align individual interests with collective welfare, fostering sustainable cooperation~\cite{ostrom2000collective}.

In the rapidly evolving intersection of technology and society, we identified a critical opportunity presented by the integration of artificial intelligence (AI) agents into socio-economic systems~\cite{rahwan2019machine}. Unlike humans, whose strategies are complex and often unpredictable, AI agents offer a malleable platform that can be precisely programmed or influenced~\cite{cath2018governing}, a realization that opens new avenues for promoting cooperation.  By controlling AI agents' propensity to cooperate, we can enhance overall system cooperation. This presents an opportunity for governments to enact rules specifically governing AI behavior~\cite{floridi2022unified}, mandating cooperative strategies and using AI as tools to promote collective interests.
As a consequence, it could be argued that the role of government extends to overseeing AI behavior in social dilemmas~\cite{taddeo2018ai}. By legislating AI conduct to favor cooperation, governments can regulate AI's impact on society and harness it to encourage cooperative behavior among all participants~\cite{hughes2018inequity}. This approach integrates technological advancements with policy interventions, aiming to create systems where cooperation is the rational and advantageous choice for all agents involved.

To illustrate how governments might interact with AI companies and their technologies, consider the fictional company ``Edison'' which manufactures self-driving cars. While these vehicles must comply with all traffic laws and regulations, there are behaviors not explicitly governed by legislation. For instance, when a driver attempts to merge onto a busy road without the right of way, self-driving cars could, within legal parameters, choose not to yield. However, Edison could program its cars to act more cooperatively, allowing others to merge smoothly without compromising safety. This added layer of courteous behavior enhances overall traffic flow and exemplifies how AI can be designed to promote cooperation beyond what regulations mandate.

Furthermore, Edison might program its cars to be especially accommodating to other vehicles of the same brand, effectively using cooperative behavior as a form of brand promotion. Such actions are currently outside the scope of governmental influence but have significant implications for public cooperation and competition. Similar scenarios exist in other AI applications: social media algorithms, AI-controlled energy grids, or any other form of AI-driven mega-infrastructure~\cite{hintze2022whose}.

\begin{figure}
    \centering
    \includegraphics[width=1.0\linewidth]{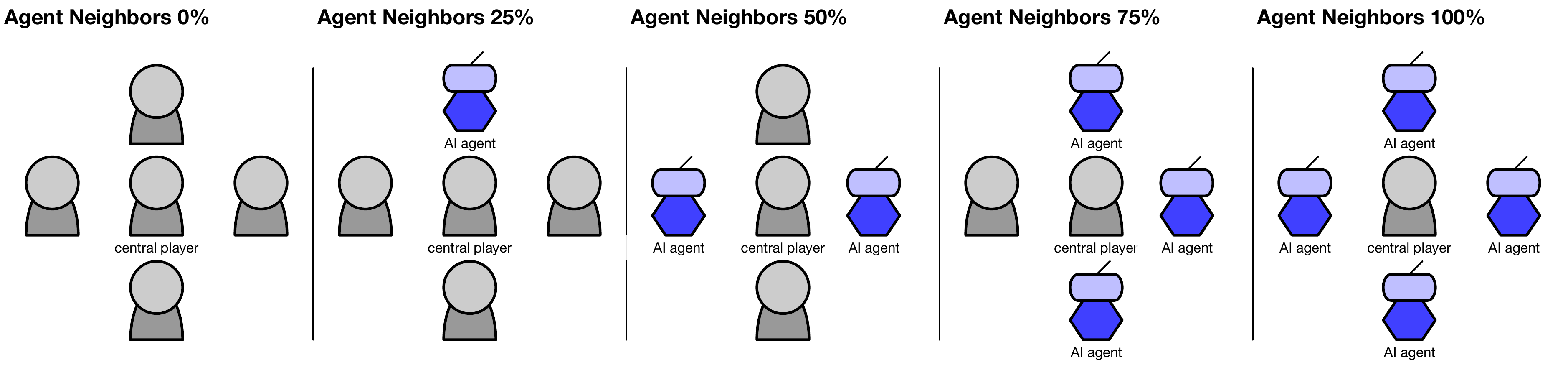}
    \caption{Illustration of AI Agent density in the neighborhood of the central player. Each panel illustrates the effect of players being replaced by AI agents, ranging from 0\% to 100\% replacement in increments of 25\% ($\rho_{A}=0.0,0.25,0.5,0.75,1.0$ respectively) -- observe that the central player is never replaced as $\rho_{A}$ only pertains to the neighborhood.}
    \label{fig:densityIllustration}
\end{figure}

The advent of AI thus opens new opportunities and raises questions about regulations or policies that promote cooperative behavior in situations akin to the public goods game. Here, we investigate different policy scenarios that influence AI agent behavior and their effectiveness in promoting cooperation. We assume a population of humans, referred to as \textit{players}, who engage in the public goods game, while a fraction of them is replaced by AI \textit{agents}. Since the extent to which AI will replace human decision-makers is uncertain, we experimentally control the density of these agents ($\rho_{A}$) (see Figure \ref{fig:densityIllustration} middle and right panel for an illustration).

Regulatory bodies could, in principle, mandate that agents behave in a certain way or allow AI companies and users deploying such agents to determine their behavior. Beyond these two extremes, we propose a third mechanism that requires agents to mimic the behavior of human players in the game -- a mimicking strategy. Human behavior is modeled here using evolutionary game theory mechanisms~\cite{adami2016evolutionary}, simplifying player behavior to a probability of cooperating or defecting. Through natural selection, players adjust their strategies based on the payoffs they receive in different situations. We will show that neither of the first two mechanisms -- forcing behavior or not regulating it -- promotes cooperation. In contrast, our proposal for agents to mimic player behavior promotes cooperation among players. Further, the more players are replaced by AI agents, the more players are encouraged to cooperate, which makes this proposal sustainable for a future in which we will find an increasing number of AI agents.

\section{Results}\label{sec3}
We will test the impact that different policies regulating AI agents that participate in the public goods game have on player behavior, using an agent-based evolutionary computational model. A population of players, each defined by a probability of cooperating in a public goods game, evolves under selection and mutation over thousands of generations. In each generation, players are grouped into small sets (of size $k+1$) to play the public goods game multiple times, from each player's perspective. Payoffs accumulate based on individual contributions and a synergy factor $r$ that amplifies the individual contributions. This amplified capital constituted the public good, which is distributed equally to all individuals in a group (see Methods for more details on the game and its implementation). 

In mathematical terms, consider a single player in a group with $k$ other players. Let $N_C$ represent the number of cooperators and $N_D$ the number of defectors among those $k$ neighbors. Then, the payoffs for the cooperator and defector can be written as
\begin{equation}
    P_C = r\frac{N_C+1}{k + 1} - 1 \label{equ:Cpayoff}
\end{equation}

\begin{equation}
   P_D = r\frac{N_C}{k + 1} \label{equ:Dpayoff}
\end{equation}
For cooperation to be favorable, the payoff for a cooperator in a group of cooperators should exceed that of a defector among defectors, that is, $P_C(N_C = k) - P_D(N_C = 0)>0 $, which implies $r > 1$.
A defector outscores a cooperator if $ P_D - P_C > 0$, which implies $r < k + 1 $.
However, for all $r$ it is favorable for all players to cooperate ($N_C=k$), which indicates that a {\em dilemma} exists for all $r$ between 1 and $k+1$~\cite{hintze2015punishment}. The parameter $r$ becomes a {\em critical parameter} controlling the game's outcome, and the point at which $ P_D - P_C = 0$ is a {\em critical} point (the relationship to the critical points of condensed matter physics is made more clear in~\cite{hintze2015punishment}). 

\begin{figure}
    \centering
    \includegraphics[width=0.8\linewidth]{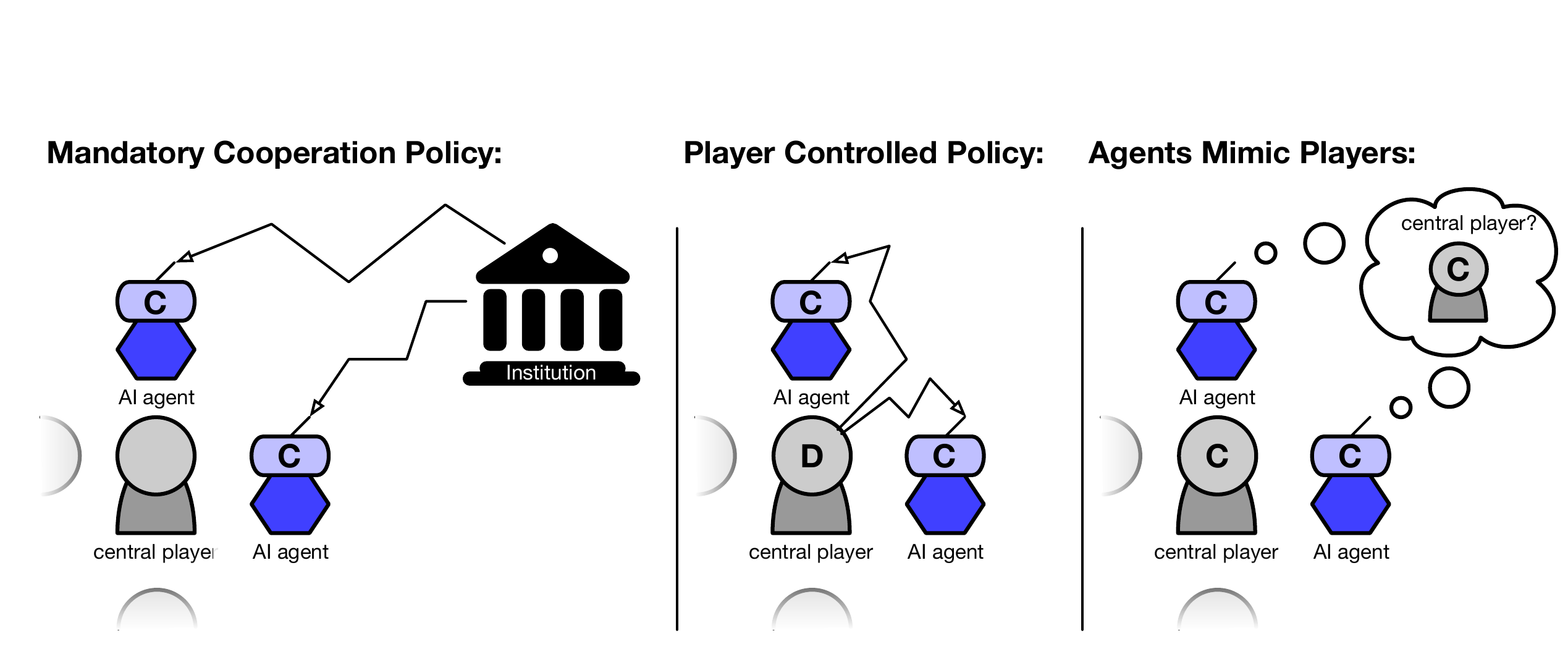}
    \caption{Illustration of three different policies for AI agent behavior. The left panel shows the ``Mandatory Cooperation Policy for Agents''; here, a hypothetical institution that mandates that all AI agents are to always cooperate. The middle panel shows the ``Player Controlled Agent Policy''; here, the central player can determine the likelihood of peripheral AI agents cooperating. However, the decision of the central player to cooperate is independent of the AI agents' policy. The right panel shows the policy where ``Agents Mimic Players''; here, the peripheral AI agents use the central players' likelihood to cooperate to determine their own choice.}
    \label{fig:policyIllustration}
\end{figure}

\subsection{Mandatory Cooperation Policy for AI Agents}
A governing body might attempt to promote cooperation by forcing AI-controlled agents to always cooperate, with the idea of ensuring that at least the agents behave well and support the public good, potentially promoting cooperation among human players (see Figure~\ref{fig:policyIllustration} left panel). However, the calculation that bounded the region in which a dilemma is present showed that the region does {\em not} depend on 
the density of cooperators, which makes it unlikely that influencing the density of cooperators will not change the boundaries and, therefore, not alleviate the dilemma. 
We experimentally confirm this prediction by creating an evolutionary scenario that implements the mandatory cooperation policy for AI agents.

\begin{figure}[H]
\centering
\includegraphics[width=\linewidth]{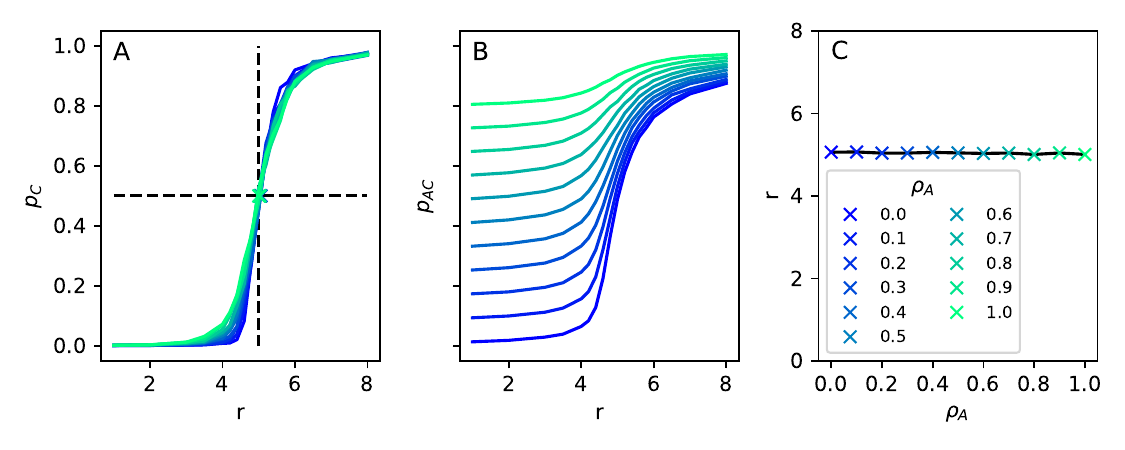}
\caption{Evolution of cooperation while cooperation of agents is mandated as a function of synergy factors $r$ A: Player probability of cooperation after evolution converges as the average of 100 independent replicates, as a function of $r$, for different AI agent densities $\rho_{A}$ indicated by the color code in panel C. B: Overall cooperation frequency within the population (players and AI agents). C: Critical point at which cooperation evolves (y-axis), given the different probabilities of encountering AI agents $(\rho_A)$ (x-axis) that are mandated to cooperate.}
\label{fig:infusion}
\end{figure}

We find the probability of cooperation after evolutionary optimization to be unaffected by the infusion of cooperating AI agents (see Figure~\ref{fig:infusion} A and C); even though the frequency of cooperating agents rises (see Figure~\ref{fig:infusion} B). As such, mandating AI agents to always cooperate only increases the cooperation with the population in proportion to the density of AI agents $\rho_{A}$, but does not encourage players to cooperate.

\subsection{Player-Controlled Agent Cooperation Policy}
In this scenario, the government does not regulate agent behavior in any way. Still, the AI companies set their agent's policies, or whoever owns or deploys an AI agent does so (see Figure~\ref{fig:policyIllustration} middle panel). To model this, each player still has a probability of cooperating ($p_{\rm C}$) that can adapt over the course of evolution to the conditions of the current scenario. In addition, each player also has at their disposal an auxiliary probability ($p_{\rm AC}$) that is used to determine the probability of an agent cooperating within its neighborhood. This auxiliary probability is also optimized by evolution. As a consequence, depending on the AI agent density, the density of cooperators around the central player is determined by the likelihood of other (non-AI agent) players and its probability reserved for agent cooperation ($p_{\rm AC}$). 

Their payoff, however, is now not only dependent on their own cooperation probability $p_{\rm C}$ but also on their auxiliary probability for agent cooperation ($p_{\rm AC}$). Evolution is optimizing both probabilities. If it is, for example, beneficial for a player to cooperate while also surrounding itself with cooperating agents, both probabilities should be maximized ($p_{\rm C}=1.0$ and $p_{\rm AC}=1.0$).

Note that the central player only pays its own ante when cooperating. In contrast, an AI agent would pay for its own cost to cooperate even though the central player controls its probability of cooperating.  As an illustration, if an AI-driven car is programmed by a driver to give way to them coming from an on-ramp,  the AI agent having to wait does not cost the driver anything.

\begin{figure}[H]
\centering
\includegraphics[width=\linewidth]{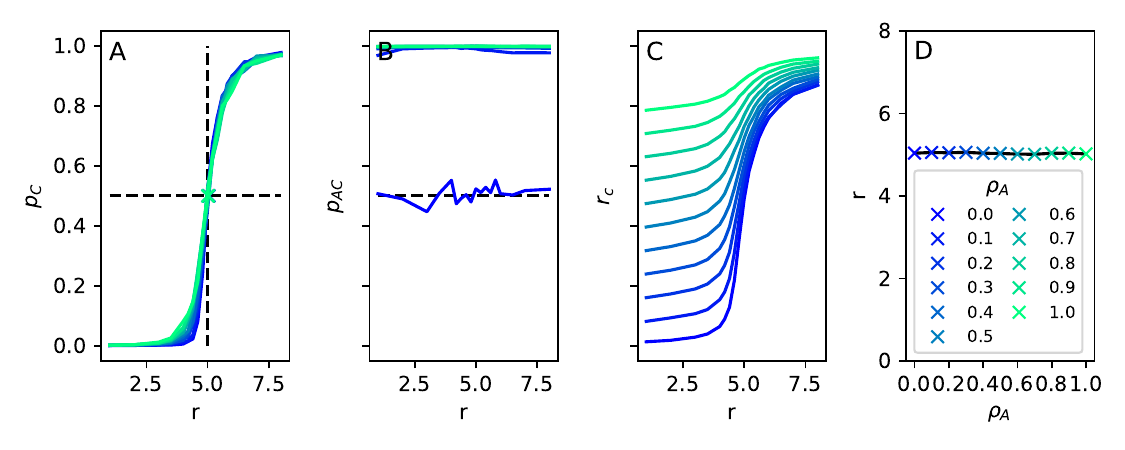}
\caption{Evolution of cooperation when agent behavior can evolve under the control of players for the same experimental setup (except the choice of actions) as for Figure \ref{fig:infusion}. A: Central player's probability to cooperate $p_{\rm C}$ at the end of evolutionary adaptation as a function of synergy parameter $r$ and AI agent density $\rho_A$ (color code in panel D). B: Evolved likelihood for an AI agent to cooperate $p_{\rm AC}$ at the end of evolution as a function of synergy $r$ and AI agent density $\rho_A$. The horizontal dashed line depicts the expected $p_{\rm AC}=0.5$ probability for drift. C: Population frequency of cooperation (all players and AI agents) as a function of $r$ and $\rho_A$. D: Critical synergy value $r$ (y-axis) at which the probability of cooperation for players exceeds $0.5$ for all experimental conditions of $\rho_{A}$ (x-axis).}
\label{fig:control}
\end{figure}

While this scenario can increase the likelihood of cooperators, we again do not expect this to reduce the synergy threshold for the player to engage in cooperation following the previous logic. We indeed find the synergy factor to be unaffected (see Figure~\ref{fig:control} A and C). While the controlled agents become cooperators the moment they appear in the neighborhood ($\rho_{A}>0.0$) as this maximizes the payoff of the player controlling them ($P_{\rm D}(N_{C})<P_{D}(N_{C}+1)$) (see Figure \ref{fig:control} B), this scenario does not affect the probability of the central player to cooperate, and thus the dilemma is not lifted. 

\subsection{Agents Mimic Players}
Having analyzed the shortcomings of both mandating cooperation and allowing players to control AI agents, we recognize the need for a fundamentally different policy. These initial strategies fall short of removing the dilemma because they do not alter the critical point at which cooperation becomes advantageous for players.  

From previous experience, we hypothesized that incentivizing players to cooperate requires increasing the payoff associated with cooperative actions. This realization led us to propose an innovative policy wherein AI agents mimic the behavior of the central player. In this paradigm, if the central player chooses to cooperate, the neighboring agents (proportional to the agent density $\rho_{A}$) will also cooperate (see Figure \ref{fig:policyIllustration}, right panel). This approach establishes a reciprocal relationship between agents and players, effectively reshaping the incentives and promoting cooperative behavior.

Although the public goods game is not an iterative game where agents can estimate the likelihood of players cooperating based on previous actions, we can model such a mechanism by defining the AI agents' cooperation probability to be identical to that of the central player. Thus, depending on the agent density $\rho_{A}$, players will encounter groups where agents mirror player behavior. If the central player is a cooperator, the likelihood of having cooperating neighbors increases proportionally to the AI agent density. This effectively enhances the expected payoff for cooperators, as they are more likely to be in groups with other cooperators.

\begin{figure}[H]
\centering
\includegraphics[width=0.6\linewidth]{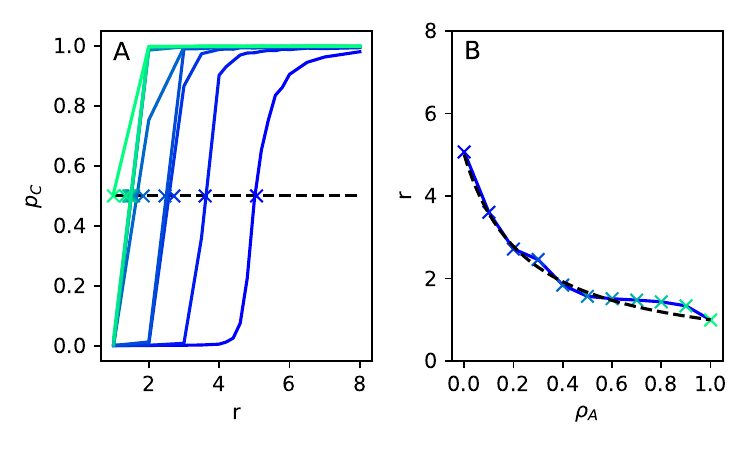}
\caption{Evolution of cooperation when agent behavior mimics player behavior, for different agent densities $\rho_{A}$, for the same experimental setup (except the choice of actions) as used for Figure \ref{fig:infusion}. A: players' probability of cooperating at the end of evolutionary adaptation $p_{\rm C}$ as a function of $r$ and AI agent density $\rho_A$ (color code in Figures \ref{fig:infusion} and \ref{fig:control}). Critical points denoted by `x'. B: Critical synergy $r$ at which the probability of cooperation for players exceeds $0.5$, obtained from crossover points in panel A marked as `x'. The dashed line represents the predicted synergy factor $r$ for cooperation to evolve given by Eq.~(\ref{equ:synergyMimic}).}
\label{fig:mimic}
\end{figure}

If $\rho_A$ is the likelihood that a peripheral player mimics the strategy of the focal player,
then if the focal player is a cooperator, the number of defectors is reduced from $N_D$ to
\begin{equation}
N_D'=(1-\rho_A)N_D\;.
\end{equation}
On the other hand, if the focal player is a defector, the number of cooperators is reduced by mimicry to
\begin{equation}
N_C'=(1-\rho_A)N_C\;.
\end{equation}
We can now calculate the cooperation condition $P_C-P_D\geq0$ using these substitutions. We find
\begin{equation}
    \frac{r(k-N_D'+1)}{k+1}-1\geq \frac{N_C'}{k+1}\;,
\end{equation}
which simplifies to
\begin{equation}
r \geq \frac{k + 1}{\rho_A \cdot k + 1}\;. \label{equ:synergyMimic}
\end{equation}
Surprisingly, this relation is independent of the number of cooperators and defectors in each neighborhood and only depends on the mimicry probability (here given by the AI agent density $\rho_A$) and the group size. It reduces to the dilemma $r\geq k+1$ in the limit $\rho_A=0$, and completely eliminates the dilemma when $\rho_A=1$.
Figure \ref{fig:mimic} shows the numerical experiments implementing this scenario and confirms the prediction Eq.~(\ref{equ:synergyMimic}). Simply put, agents that mimic player behavior reduce the synergy threshold $r$ so that cooperation can evolve at much-reduced synergies, promoting collaboration within the group and thus changing player behavior.

\section{Discussion}\label{sec4}
We explored the impact of different policy scenarios on promoting cooperation in the public goods game, specifically in the context of integrating AI agents into human social systems. Our findings highlight the nuanced role that AI agents can play in influencing human cooperative behavior, depending on how their actions are regulated or programmed.

We first examined the \textbf{Mandatory Cooperation Policy for AI Agents}, where AI agents are forced institutionally to always cooperate. Contrary to intuitive expectations, this policy did not enhance cooperation among human players. While the overall frequency of cooperative actions increased due to the agents' enforced cooperation, the critical threshold for human players to find cooperation advantageous remained unchanged. This outcome aligns with theoretical predictions that the critical point in the public goods game is independent of the frequency of cooperators in a player's immediate environment~\cite{hardin1968tragedy,hintze2015punishment}. The inability of this policy to shift the cooperative dynamics (and thus lift the dilemma) suggests that simply injecting cooperative agents into the system is insufficient to influence human behavior in the desired manner.

Next, we investigated the \textbf{Player-Controlled Agent Cooperation Policy}, where human players determine the cooperation probability of AI agents in their vicinity. Despite the potential for players to benefit from surrounding themselves with cooperative agents, the evolutionarily optimized strategies did not lead to increased cooperation among players themselves. Players evolved to maximize their own payoffs by programming agents to cooperate while choosing to defect themselves, exploiting the cooperative agents without reciprocating. This mirrors real-world scenarios where individuals might leverage AI systems for personal gain without contributing to the collective good, highlighting a potential pitfall in allowing unrestricted control over AI behavior.

In contrast, the \textbf{Agents Mimic Players} policy yielded a significant shift in cooperative dynamics and was successful in removing the dilemma. By programming AI agents to mirror the cooperation probability of the central player, we introduced a mechanism akin to conditional cooperation or reciprocity~\cite{nowak2006five}. This policy effectively lowered the synergy threshold ($r$) required for cooperation to be advantageous for players (see Equation~\ref{equ:synergyMimic}). As AI agents copy the players' strategies, cooperative players find themselves in increasingly cooperative groups, enhancing their payoffs and promoting the evolution of cooperation. This outcome underscores the power of reciprocity and conditional strategies in fostering cooperative behavior~\cite{fehr2002altruistic,ostrom2000collective}.

Note that this mimicking strategy could also have AI agents copying the defective behavior of players. However, the mimicking policy of AI agents shifts the critical synergy threshold so that it becomes more lucrative for the players to cooperate. Thus, the ``selfish'' thing to do in this scenario is to cooperate.

The success of the mimicry policy suggests that AI agents can serve as catalysts for cooperation when their behavior is contingent upon that of human players. By aligning agents' actions with players' choices, we create a feedback loop that rewards cooperative behavior and penalizes defection. This mechanism leverages principles from evolutionary game theory, where strategies that promote mutual cooperation can become stable under certain conditions~\cite{nowak2006five,adami2016evolutionary}. While the mechanism that creates mimicry is not specified, we can imagine that AI agents use information gathered from observing player actions to mimic the players, thus establishing again that it is information that promotes cooperation~\cite{Iliopoulosetal2010}.

\subsection{Implications for Policy and Governance}
Our findings offer vital insights for policymakers and institutions aiming to foster cooperative behavior in an era increasingly dominated by AI technologies. By introducing and validating a novel policy where AI agents mimic human behavior, we provide a transformative framework that could redefine strategies for enhancing cooperation in complex social systems. Mandating cooperation from AI agents or allowing users to control agent behavior without oversight may not lead to the desired increase in human cooperation. Instead, policies that encourage AI systems to respond adaptively to human behavior --specifically by mimicking or reciprocating actions -- could foster more cooperative interactions.

The approach outlined here aligns with concepts of responsive regulation and adaptive governance, where rules and systems are designed to be flexible and responsive to the behaviors of individuals within the system~\cite{ostrom1990governing,floridi2022unified}. By integrating AI agents that promote reciprocity, governments and institutions can harness technology to enhance social welfare without imposing rigid mandates that may be ineffective or counterproductive.

\subsection{Limitations}
The use of evolutionary game theory (EGT) as the framework for modeling human behavior has important limitations. EGT assumes that strategies evolve over time based on their relative success, akin to natural selection in biological systems. However, human decision-making is influenced by a myriad of factors beyond simple payoff maximization, including emotions, social norms, and cognitive biases~\cite{camerer2011behavioral}. Consequently, the extent to which EGT can accurately capture the complexities of human cooperation is debatable. While EGT has been successfully applied in economics and biology to model strategic interactions~\cite{smith1982evolution}, its assumptions may oversimplify the nuanced processes underlying human behavior.

Moreover, the agents in our simulations are controlled via parameters that directly influence their likelihood to cooperate or defect. In scenarios like ``Mandatory Cooperation'' or ''Player-Controlled Agents,'' we impose external rules or allow players to evolve control over agents, respectively. However, in real-world settings, the mechanisms by which human behavior would determine agent actions are not straightforward. Autonomous agents, particularly those powered by advanced AIs such as large language models, operate on the basis of complex algorithms and vast datasets~\cite{rafailov2024direct}. Directly coupling player behavior to agent actions, as done in our models, may not accurately reflect how such systems function in practice. However, governmental rulings on AI policies can pave the path to integrate AI agents such that they promote cooperation in the ways exemplified here. 

To address the discrepancy between models and reality, future research could explore more sophisticated models where agent behavior emerges from interactions with players rather than being directly controlled. For instance, incorporating reinforcement learning mechanisms could allow agents to adapt their strategies based on observed player actions over time~\cite{sutton2018reinforcement}. This approach would better mirror the iterative and dynamic nature of human-agent interactions, capturing feedback loops and mutual adaptation. It also presents an opportunity to determine agent behavior. As mentioned before, reinforcement learning based on the heuristic to \textit{do upon others as done upon yourself} can implement the mimicking strategy for AI agents, thus promoting cooperation among humans.

Another methodological limitation lies in the assumption of well-mixed populations and the absence of spatial structure in our simulations. In reality, interactions often occur within networks where the structure can significantly impact the evolution of cooperation ~\cite{santos2006evolutionary,SzaboFath2007}. Incorporating network topology into the model could provide a more accurate representation of social dynamics, potentially altering the outcomes observed in our study. However, we remark that spatial proximity in itself tends to promote cooperation, so we do not expect structured populations to negate our findings.

We note that the ``Agents Mimic Players'' scenario presupposes that agents can perfectly observe and replicate player behavior without error. In practical applications, noise and imperfect information are inevitable~\cite{fudenberg2009learning}. Introducing stochastic elements or uncertainties into the agents' observation and imitation processes could yield more realistic results, shedding light on how robust the cooperative dynamics are to such imperfections. In addition, our model simplifies the cost and benefit structure of cooperation and defection to a single parameter, the synergy factor $r$. While this abstraction is useful for theoretical exploration, real-world public goods scenarios involve more complex and often non-linear cost-benefit relationships ~\cite{ostrom1990governing}. Extending the model to include variable costs, diminishing returns, or threshold effects could enhance its applicability to real situations.

Finally, ethical considerations regarding the deployment of autonomous agents that can influence human behavior should not be overlooked. The implementation of agents designed to promote cooperation raises questions about manipulation, consent, and the potential for unintended consequences~\cite{bryson2018patiency}. However, promoting cooperation is in the interest of all and thus might be easily accepted. Further, ethical AI guidelines generally seek to prevent malicious or deceptive behavior and usually explicitly try to make AI behave ``well'', which mimicry encourages as long as the player they are emulating behaves well.

\section{Conclusion}\label{sec5}
This study examined the potential of artificially intelligent agents to promote cooperation in public goods games by testing three different scenarios: \textit{Mandatory Cooperation Policy for AI Agents}, \textit{Player-Controlled Agent Cooperation Policy}, and \textit{Agents Mimic Players}. Our computational simulations revealed that only when agents mimic player behavior does the critical synergy threshold $r$ for the emergence of cooperation decrease, thus lifting the dilemma. This finding suggests that agents designed to reciprocate player actions can foster cooperative behavior among humans.

While our model simplifies complex real-world interactions, it provides a theoretical foundation for how AI agents might be leveraged to enhance cooperation in societal dilemmas. The assumption that agents can perfectly mimic player behavior is an idealization; in practice, implementing such strategies may involve challenges related to imperfect information and stochasticity. Future research should explore more sophisticated models that incorporate learning mechanisms, uncertainties, and network structures to better reflect the intricacies of human-agent interactions.

Ethical considerations are paramount when deploying AI agents that can influence human behavior. Ensuring that such agents promote cooperation without infringing on individual autonomy or causing unintended consequences is crucial. As AI continues to integrate into various aspects of daily life, understanding and guiding its impact on social dynamics becomes increasingly important.

In conclusion, our study illuminates a pivotal role for AI agents as catalysts for cooperation in social dilemmas. By uncovering and validating a novel mechanism wherein agents mimic player behavior, we offer a new paradigm that could profoundly impact how cooperative behavior is fostered in increasingly complex societal interactions. By designing agents that reciprocate human actions, we can create environments that encourage cooperative behavior, ultimately contributing to more harmonious and efficient societies.

\section{Methods}\label{sec2}
\subsection{Computational Evolutionary Model}
We model the public goods game using evolutionary game theory. Each player's behavior is characterized by an individual probability to cooperate ($p_{\rm C}$) in a group of $k$ players (see Figure 1, here we use $k=4$). The player population is initialized with random strategies ($p_{\rm C}=0.5$). The score that each player accumulates by playing multiple ($k+1$) games with others in a group in each generation determines their likelihood of propagating to the next generation via Roulette Wheel selection~\cite{Goldberg1989}. At each generation, a player's cooperation probability may mutate (with rate $\mu=0.01$). In that case, a new $p_{\rm C}$ is drawn from a uniform distribution $[0.0,1.0]$, which will determine how likely the player will cooperate in the next generation. We use a population of 1,024 players arranged on a $32 \times 32$ grid, which allows us to control the number of interacting neighbors more easily. However, players are replaced independently of their location at every generation, ensuring a well-mixed population.

In each round, some players could be replaced by AI agents and would thus miss out on possibly receiving a payoff. Therefore, each player engages in five games as the ``central'' player together with $k$ other players or agents. In these games, only the score for the central player accumulates to avoid imbalances due to missed games among the neighbors. The peripheral players or agents receive no payoff until it is their turn to be the central player.

Experiments are run for 10,000 generations, and the line of descent (LOD), as well as the number of actions (C or D) taken by the entire population in each generation are recorded. The LOD connects a random player in the last generation through its direct ancestors back to a player in the initial population~\cite{adami2016evolutionary}. Every mutation that fixed in the population (as it led to a selective sweep) will be on this line, which tells the tale of this particular evolutionary history. Since the LOD also contains the most recent common ancestor (MRCA, which is relatively close to the end of the line) we always ignore the last 25\% of the LOD in our analyses so as only to include the history beyond the MRCA of the population.  

In one scenario where players are allowed to control AI agents, players possess a second probability that also undergoes evolutionary adaptation. The only difference is that this second probability controls AI agents rather than the player's own behavior.

\backmatter


\bmhead{Acknowledgements}

This work was supported in part through computational resources and services provided by the Institute for Cyber-Enabled Research at Michigan State University.



\begin{itemize}
\item Funding\\
Not applicable
\item Conflict of interest/Competing interests\\ 
The authors declare no conflict of interest.
\item Ethics approval and consent to participate\\
Not applicable
\item Consent for publication\\
Not applicable
\item Data availability \\
Not applicable
\item Materials availability\\
Not applicable
\item Code availability \\
The code will be made available on OpenScienceFramework.
\item Author contribution\\
A.H. conceived the research idea and experimental framework, designed the experiments and developed the methodology, performed the experiments, and analyzed the data. C.A. was responsible for formal analysis. A.H. and C.A. contributed to writing the manuscript by drafting and revising it. Both authors supervised and coordinated the research work. All authors reviewed the manuscript.
\end{itemize}

\noindent



\end{document}